\documentclass[preprint,5p]{elsarticle}
\usepackage{amsmath}
\usepackage{subfig}
\usepackage{graphicx}
\usepackage{stackengine}
\usepackage[export]{adjustbox}
\usepackage{units}
\usepackage[version=4]{mhchem}
\usepackage{makecell}

\makeatletter
\def\ps@pprintTitle{
  \let\@oddhead\@empty
  \let\@evenhead\@empty
  \def\@oddfoot{\footnotesize\itshape\hfill\today}
  \let\@evenfoot\@oddfoot
}
\makeatother

\begin{document}

\title{Size-dependent phase morphologies in \ce{LiFePO4} battery particles}
\author[]{Daniel A. Cogswell\fnref{chemE}}
\ead{cogswell@alum.mit.edu}

\author[]{Martin Z. Bazant\corref{cor1}\fnref{chemE,math}}
\ead{bazant@mit.edu}

\cortext[cor1]{Corresponding author}
\address[chemE]{Department  of  Chemical  Engineering, Massachusetts  Institute  of  Technology,  Cambridge,  MA  02139  USA}
\address[math]{Department  of  Mathematics, Massachusetts  Institute  of  Technology,  Cambridge,  MA  02139  USA}
\date{\today}

\begin{abstract}
Lithium iron phosphate (\ce{LiFePO4}) is the prototypical two-phase battery material, whose complex patterns of lithium ion intercalation provide a testing ground for theories of electrochemical thermodynamics. Using a depth-averaged (a-b plane) phase-field model of coherent phase separation driven by Faradaic reactions, we reconcile conflicting experimental observations of diamond-like phase patterns in micron-sized platelets and surface-controlled patterns in nanoparticles. Elastic analysis predicts this morphological transition for particles whose a-axis dimension exceeds the bulk elastic stripe period. We also simulate a rich variety of non-equilibrium patterns, influenced by size-dependent spinodal points and electro-autocatalytic control of thermodynamic stability. 
\end{abstract}


\maketitle

\section{Introduction}
The morphology of two-phase coexistence in single crystals of the battery material \ce{LiFePO4} has been the subject of much debate since its discovery over two decades ago \cite{Padhi1997}. Originally thought to be a low-rate material due the presence of phase boundaries between \ce{FePO4} and \ce{LiFePO4}, the material is now routinely used for high-rate applications \cite{Nitta2015}. At the same time, particle sizes have been reduced to the nanoscale, fueling interest in size-dependent and rate-dependent phase morphologies, which broadly affect intercalation-based electrochemical devices~\cite{Li2018review}. 

Such a dramatic reversal of fortune has drawn attention to this system as a model for studying the role of phase-separation in electrochemical systems, and spurred the development of sophisticated imaging techniques designed to image the phase state of single crystals. Observations of the morphology include stripes \cite{Chen2006, Ramana2009, Ohmer2015,Zhang2015}, lithiated cores \cite{Boesenberg2013, Shapiro2014,Nakamura2014,Wang2014,Lucas2015,Yu2015,Wang2016,Lachal2017}, delithiated cores \cite{Laffont2006}, both lithiated and delithiated cores \cite{Honda2015}, complex nonequilibrium morphologies \cite{Li2015,Mu2016,Lim2016,Li2018surface}, and mosaic patterns of lithiated and delithiated particles (i.e. only single-phase particles) \cite{Chueh2013,Li2014}. The reason why many different morphologies occur in one system, however, has remained a puzzle. 

Here, with the help of phase-field modeling, we show that elastic strain energy leads to significant morphological differences between nano and micro-sized particles. Whereas the equilibrium morphology of nanoparticles is controlled by elastic interaction between particle surfaces, that of microparticles is controlled by bulk elasticity. There is a stress-induced transition that depends on the size and aspect ratio of the particle, around \unit[250]{nm}  for standard platelet \ce{LiFePO4} particles, and the size-dependent nonequilbirium morphology is further influenced by driven intercalation reactions.

\begin{figure}
 \includegraphics[width=\columnwidth]{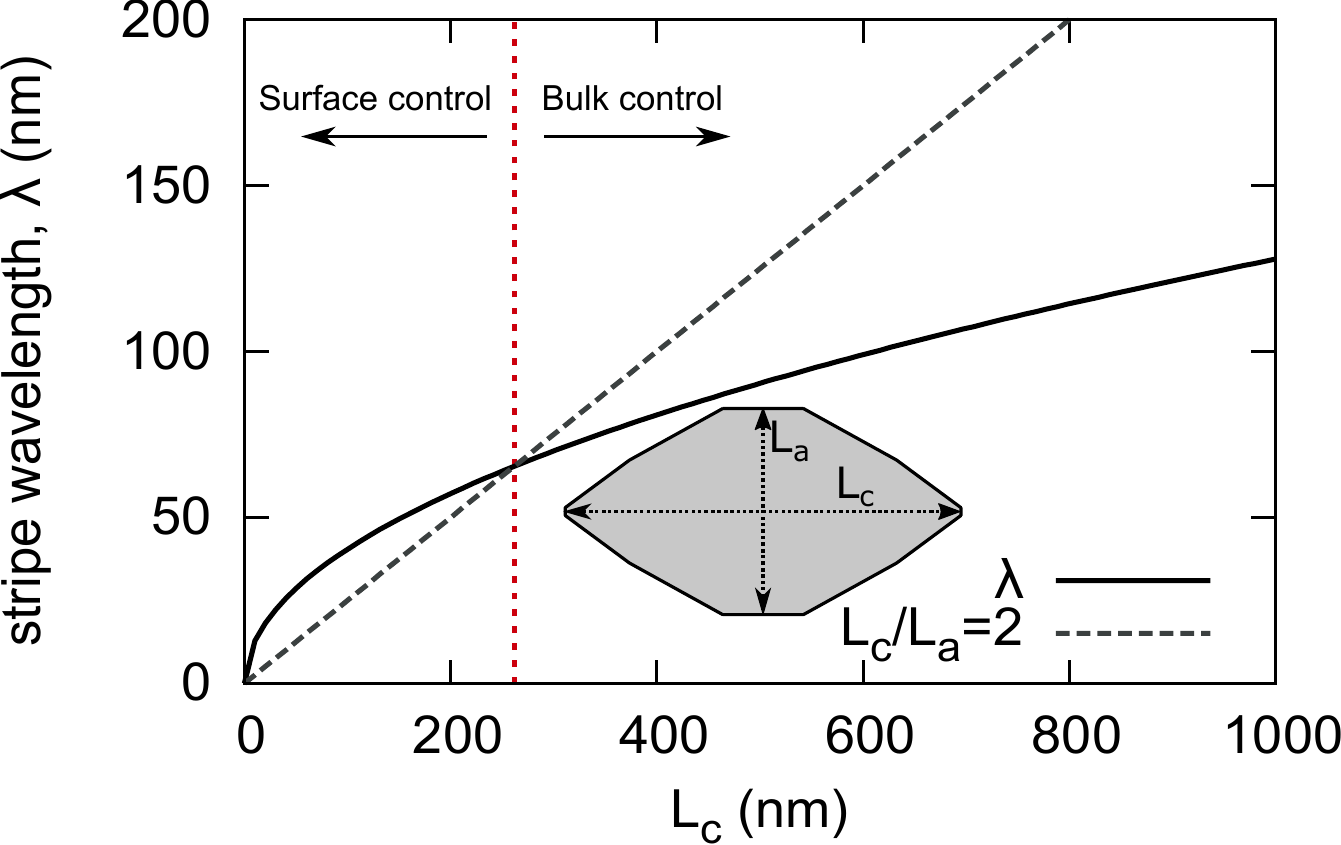}
 \caption{The transition from nano- to micro-phase morphology in \ce{LiFePO4} depends on the particle aspect ratio and occurs when the elastic stripe period exceeds half of the particle size along the a-axis.}
 \label{Fig:stripesize}
\end{figure}

\section{Theory}
Previously, we showed that equilibrium phase separation in finite-size particles involves stripes, whose wavelength scales with the square root of particle size $L_c$ along the c-axis \cite{Cogswell2012}:
\begin{equation}
 \lambda=\sqrt{\frac{2\gamma L_c}{\Delta f}}
 \label{Eq:period}
\end{equation}
where $\gamma$ is the \ce{FePO4}/\ce{LiFePO4} interfacial energy and $\Delta f$ is the free energy difference between the homogeneous and the elastically phase-separated states. Equation \ref{Eq:period} is plotted in Fig. \ref{Fig:stripesize} using the parameter values from ref. \cite{Cogswell2012}, obtained by fitting solubility data to experimentally observed stripe periods.

Although the stripe period is determined by $L_c$, the size of the particle along the c-axis, the size $L_a$ along the a-axis determines whether the stripes fit in the particle. Thus the aspect ratio $L_c/L_a$ plays an important role in determining phase boundary morphology.
In \cite{Cogswell2012} we observed that the phases tend to form a 3-layer structure as the critical particle size is approached, and that the critical particle size is $L_a=2\lambda$. Here we consider a typical \ce{LiFePO4} particle with an aspect ratio of $L_a=L_c/2$, illustrated in Fig. \ref{Fig:stripesize}. The corresponding stripe period is plotted in Fig. \ref{Fig:stripesize}, and the intersection of the curves is the smallest particle size at which this 3-layer structure can form. To the left of the intersection, the elastic stripe size too big for the particle, which is controlled by surface properties. This size regime was the focus of our previous work \cite{Cogswell2013}. However stripes become possible in particles larger than the intersection, leading to complex morphologies in large particles which we will now investigate.

\begin{figure*}[!p]
\centering
\captionsetup[subfloat]{farskip=5ex}

\subfloat[150\,nm particles at X=.5]{
\begin{tabular}{ccccccc}
  & & & $i/i_0$ & &\\
\hline
 & 0.1 & 0.05 & 0.02 & 0.01 & 0.001\\
\hline\\
 \makecell{delithiation\\(charge)}
    & \includegraphics[valign=m,width=.15\columnwidth]{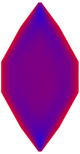}
    & \includegraphics[valign=m,width=.15\columnwidth]{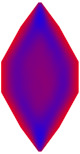}
    & \includegraphics[valign=m,width=.15\columnwidth]{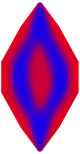}
    & \includegraphics[valign=m,width=.15\columnwidth]{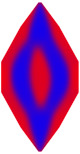}
    & \includegraphics[valign=m,width=.15\columnwidth]{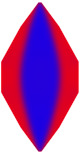}\\ 
 \makecell{lithiation\\(discharge)}
    & \includegraphics[valign=m,width=.15\columnwidth]{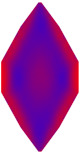}
    & \includegraphics[valign=m,width=.15\columnwidth]{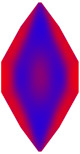}
    & \includegraphics[valign=m,width=.15\columnwidth]{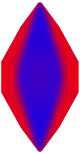}
    & \includegraphics[valign=m,width=.15\columnwidth]{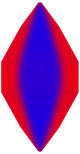}
    & \includegraphics[valign=m,width=.15\columnwidth]{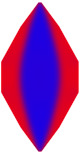}
\end{tabular}

\begin{tabular}{c}
\\
\\
\\
\includegraphics[valign=t,width=1.3cm]{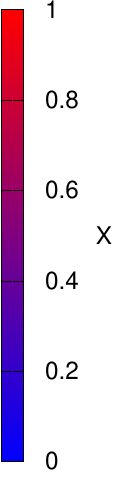}
\end{tabular}

\label{Fig:150nm}}

\subfloat[1\,um particles at X=.5]{
\begin{tabular}{cccccc}
  & & & $i/i_0$ & & \\
\hline
 & 0.1 & 0.05 & 0.02 & 0.01 & 0.001 \\
\hline\\
 \makecell{delithiation\\(charge)}
    & \includegraphics[valign=m,width=.15\columnwidth]{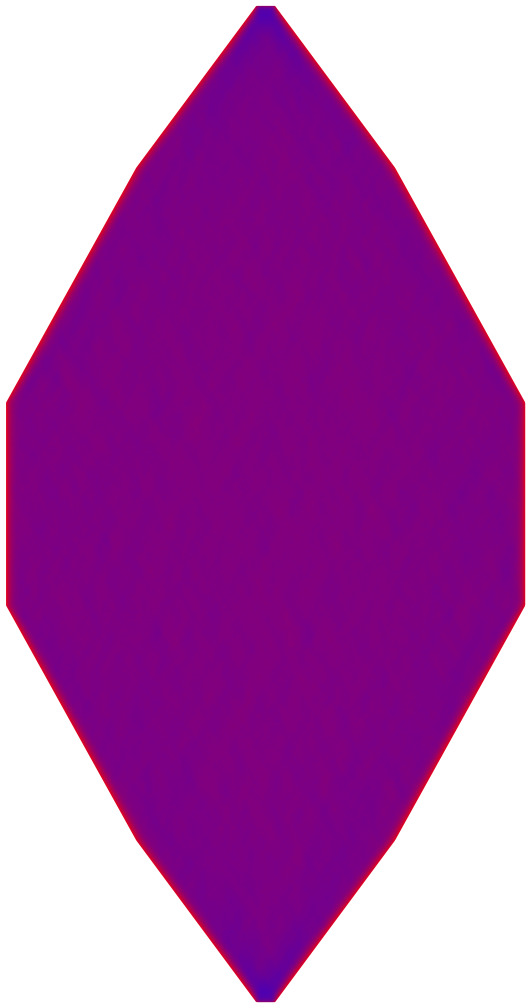} 
    & \includegraphics[valign=m,width=.15\columnwidth]{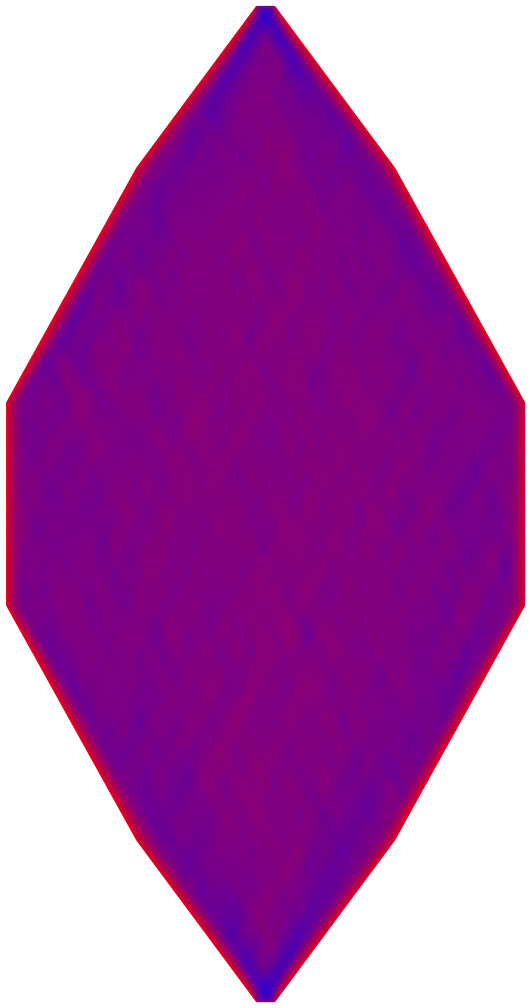} 
    & \includegraphics[valign=m,width=.15\columnwidth]{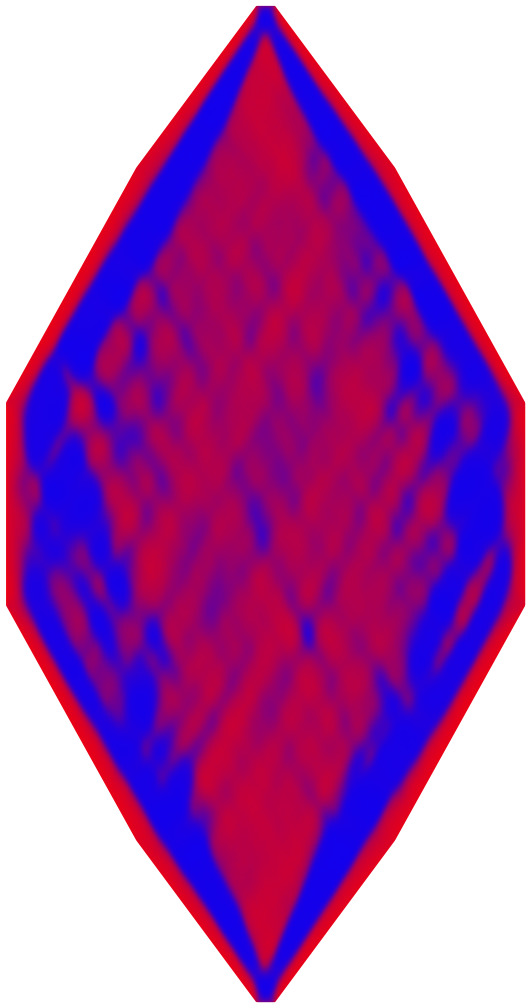}
    & \includegraphics[valign=m,width=.15\columnwidth]{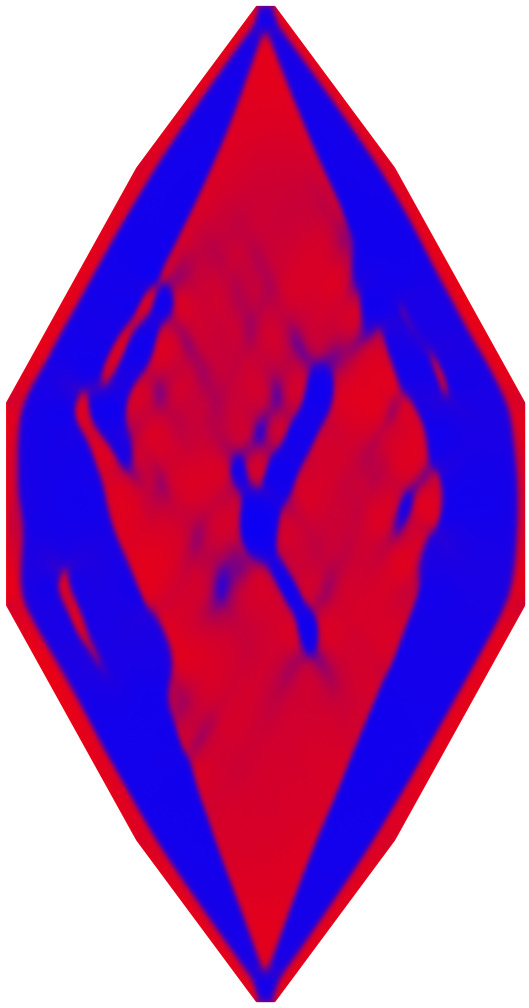}
    & \includegraphics[valign=m,width=.15\columnwidth]{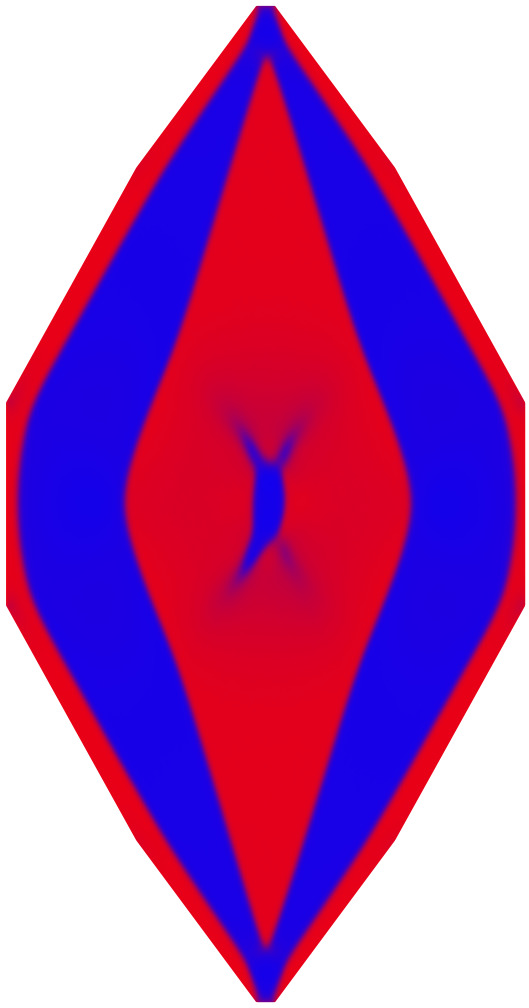} \\ 
 \makecell{lithiation\\(discharge)}
    & \includegraphics[valign=m,width=.15\columnwidth]{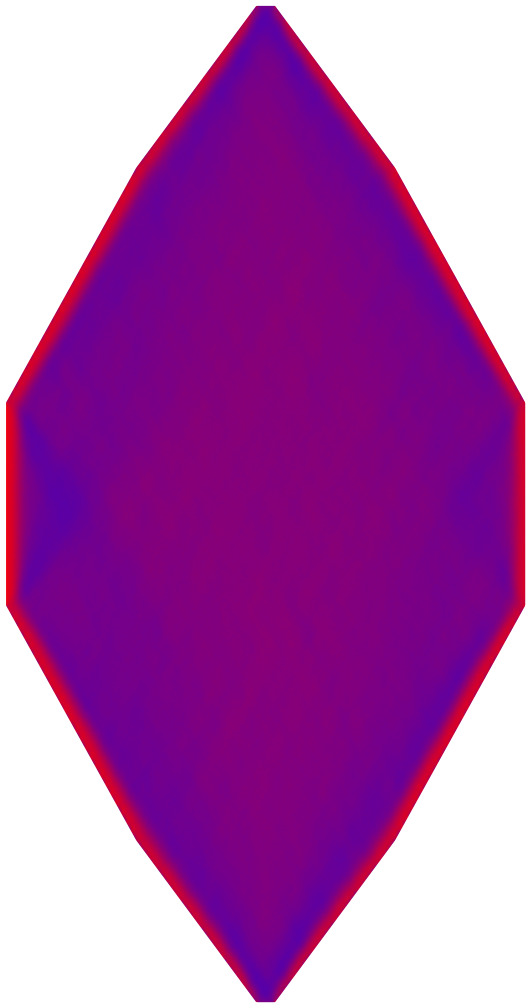}
    & \includegraphics[valign=m,width=.15\columnwidth]{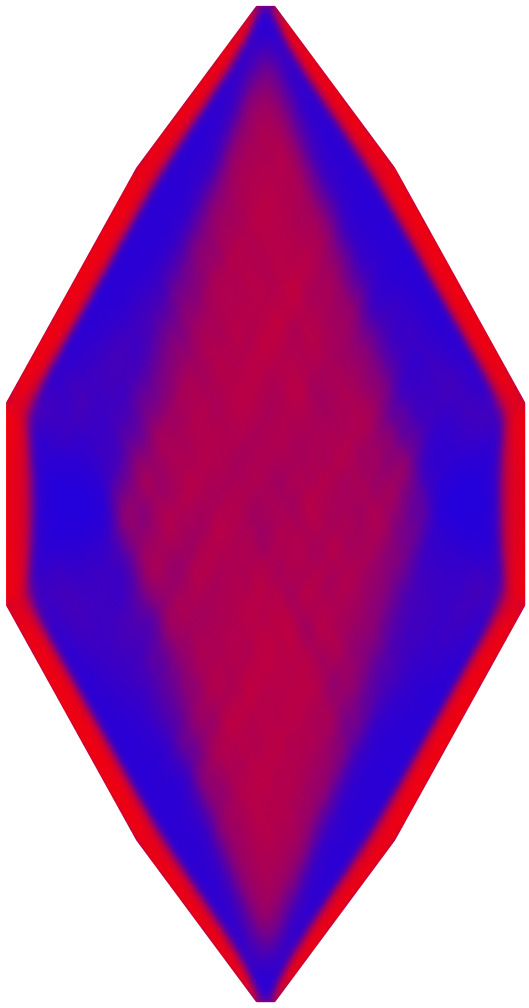}
    & \includegraphics[valign=m,width=.15\columnwidth]{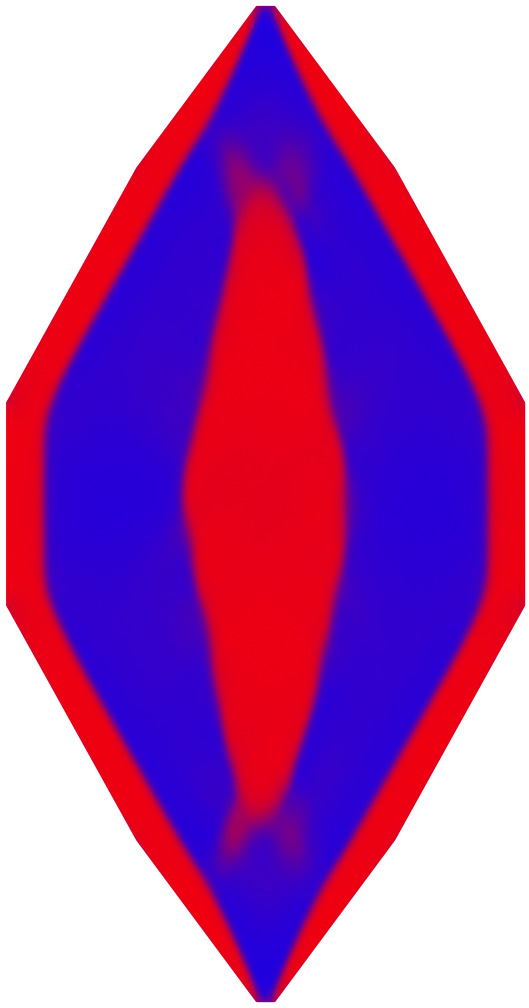}
    & \includegraphics[valign=m,width=.15\columnwidth]{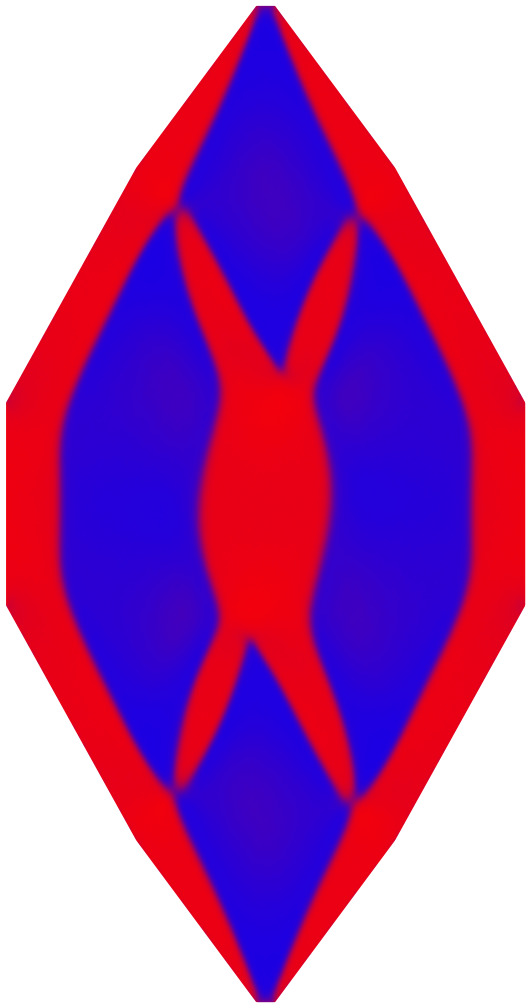}
    & \includegraphics[valign=m,width=.15\columnwidth]{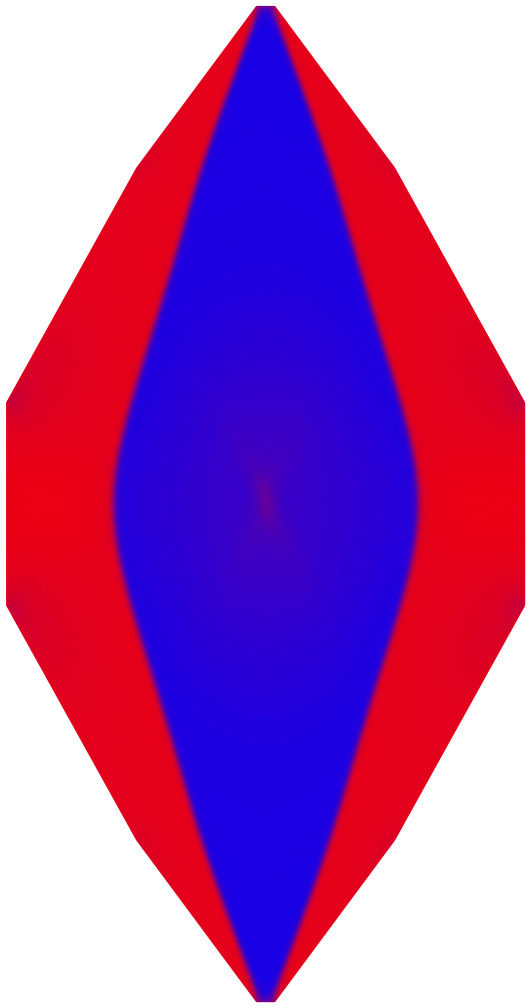} 
\end{tabular}
\hspace{1.3cm}
\label{Fig:1um}}

\subfloat[2\,um particles at X=.5]{
\begin{tabular}{cccccc}
  & & & $i/i_0$ & & \\
\hline
 & 0.1 & 0.05 & 0.02 & 0.01 & 0.001 \\
\hline\\
 \makecell{delithiation\\(charge)}
    & \includegraphics[valign=m,width=.15\columnwidth]{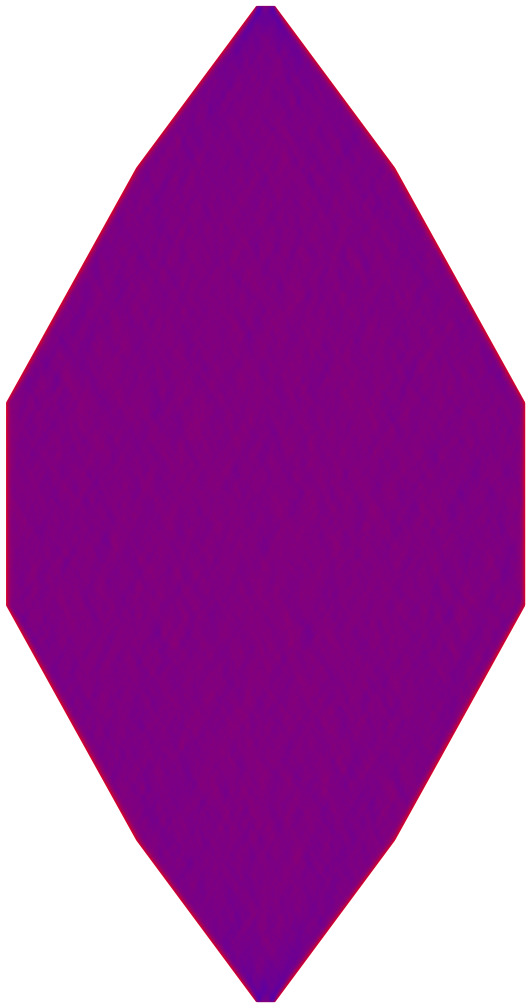} 
    & \includegraphics[valign=m,width=.15\columnwidth]{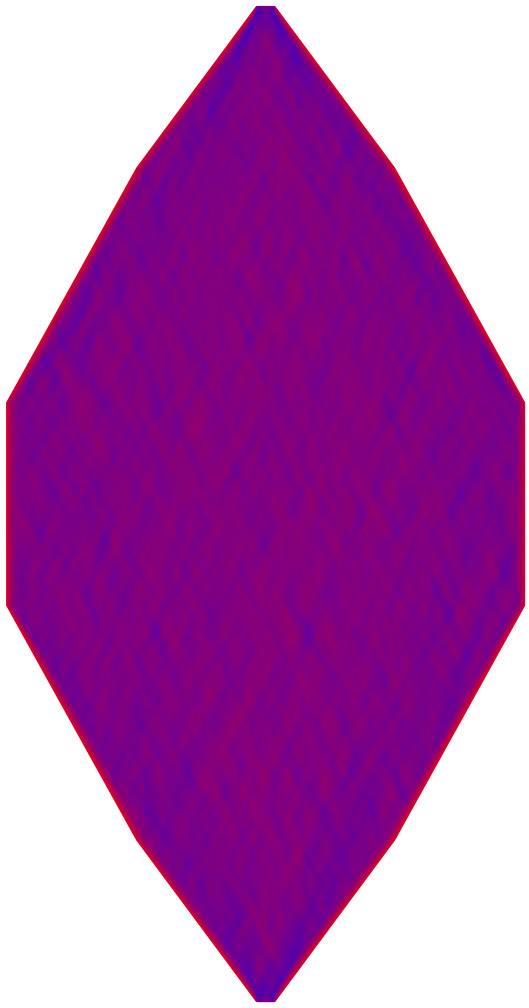} 
    & \includegraphics[valign=m,width=.15\columnwidth]{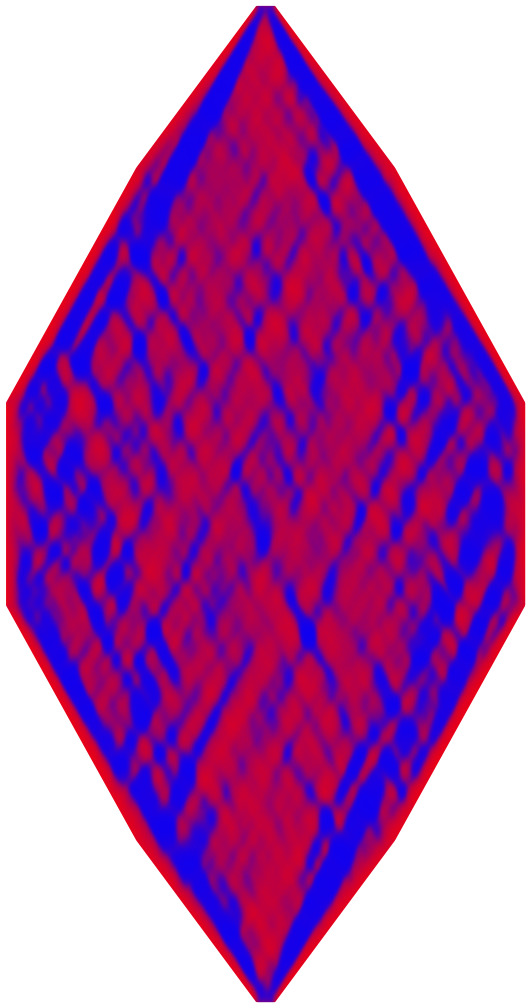}
    & \includegraphics[valign=m,width=.15\columnwidth]{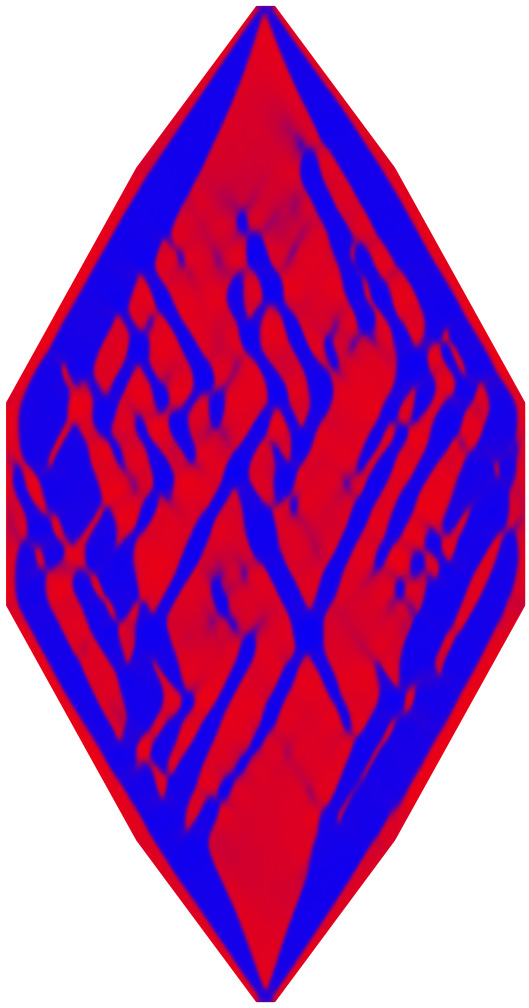}
    & \includegraphics[valign=m,width=.15\columnwidth]{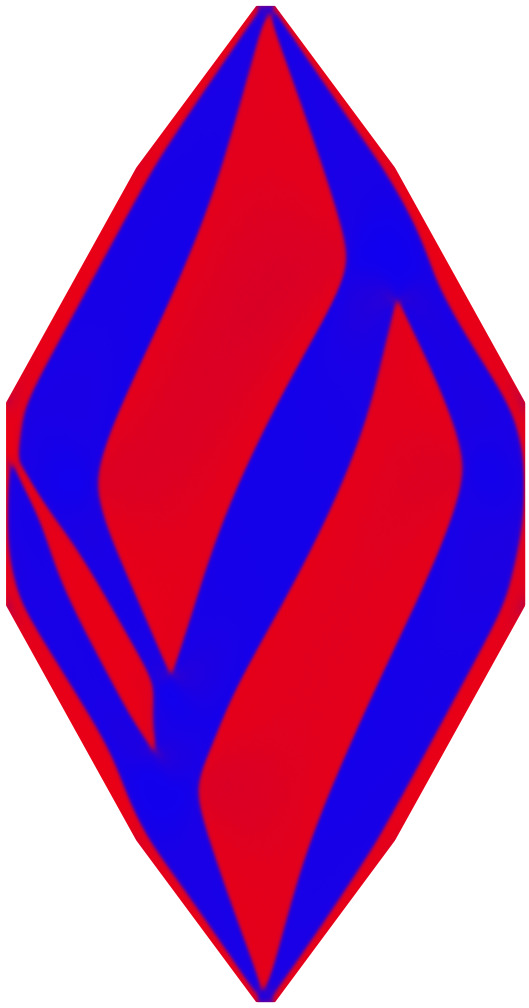} \\ 
 \makecell{lithiation\\(discharge)}
    & \includegraphics[valign=m,width=.15\columnwidth]{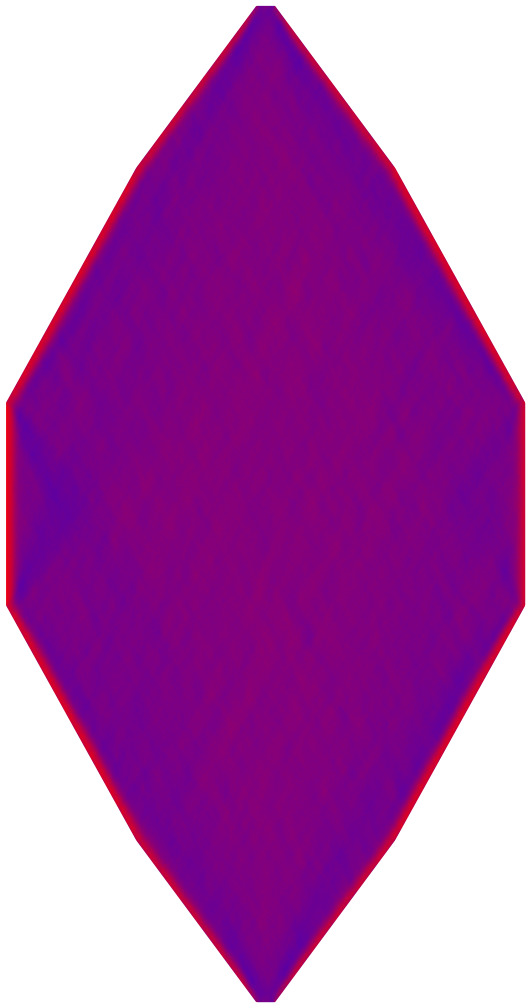}
    & \includegraphics[valign=m,width=.15\columnwidth]{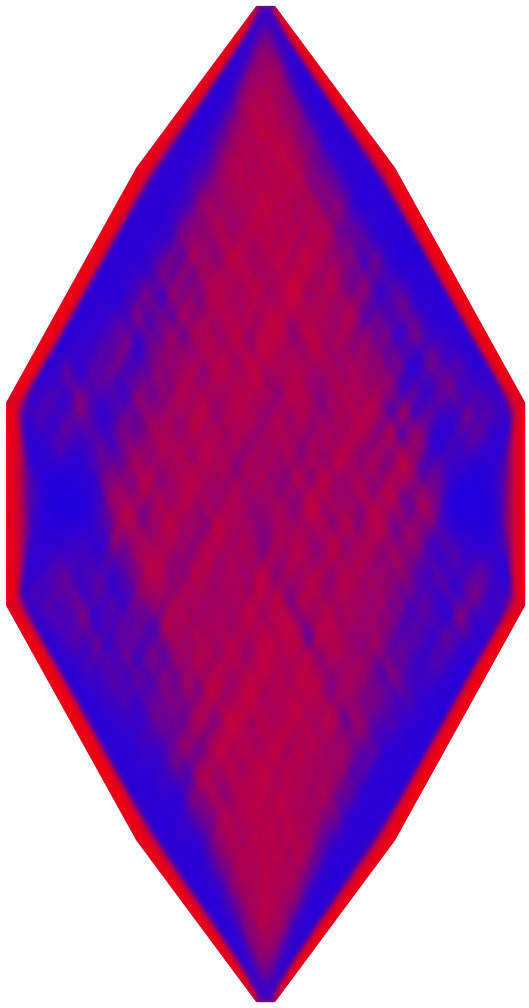}
    & \includegraphics[valign=m,width=.15\columnwidth]{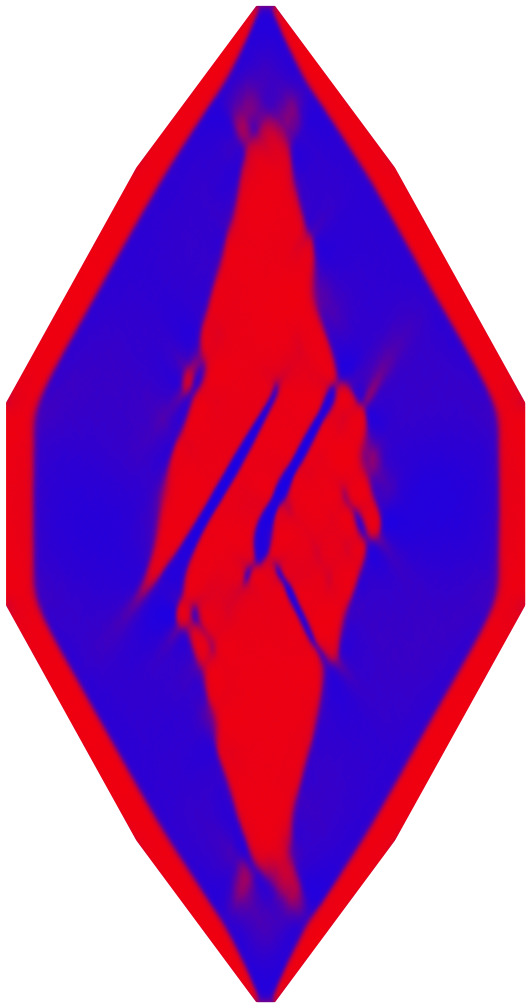}
    & \includegraphics[valign=m,width=.15\columnwidth]{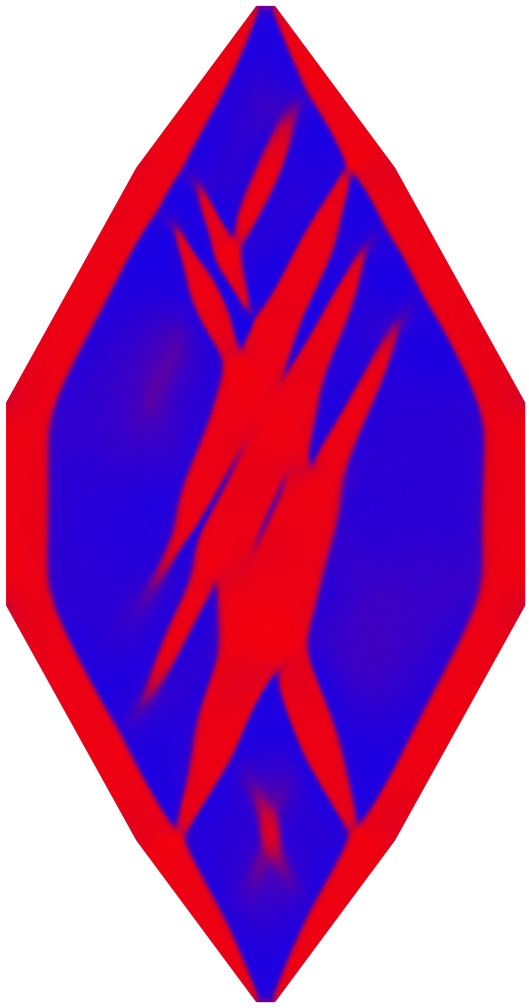}
    & \includegraphics[valign=m,width=.15\columnwidth]{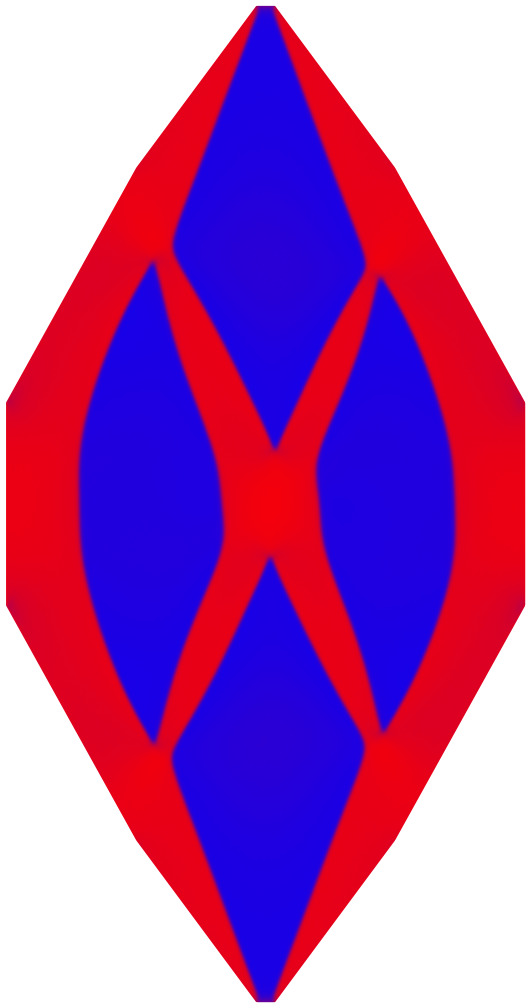}
\end{tabular}
\hspace{1.3cm}
\label{Fig:2um}}
\caption{Nonequilibrium lithium concentration in \ce{LiFePO4} particles at X=.5. Comparison of lithiation and delithiation in (a) \unit[150]{nm} (b) \unit[1]{um} and (c) \unit[2]{um} particles. The \ce{LiFePO4} phase is red, and the \ce{FePO4} phase is blue. $i/i_0$ is the ratio of the applied current to the exchange current.}
\label{Fig:particles}
\end{figure*}

\section{Results and Discussion}

We perform simulations using the same depth-averaged reaction-limited model~\cite{Singh2008,Bai2011,Bazant2013} from our previous work \cite{Cogswell2012,Cogswell2013}, but with significantly larger particles. While the generalized Butler-Volmer reaction model of this work accurately predicts near-equilibrium structures, patterns that form far from equilibrium are sensitive to the form of the reaction rate model \cite{Bazant2013,Bai2014,Bazant2017}, which is beyond our scope to investigate here. The reaction limitation assumption has recently validated by combined \textit{operando} microscopy and phase-field simulation work \cite{Hong2017}. The depth-averaged approximation has proven reliable for \ce{LiFePO4} due to the fact that elasticity promotes phase boundary alignment in the (010) plane \cite{Cogswell2012,Ohmer2015}, and the (010) surfaces where insertion occurs are quickly replenished with lithium \cite{Kobayashi2016}. A much larger interfacial width in the [010] direction was also recently shown to help validate the depth-averaged approximation \cite{Nadkarni2018}.

Fig. \ref{Fig:particles} shows simulated particles of different sizes lithiated and delithiated at different rates to 50\% SOC. 
The transition from nano to micro is visible in Fig. \ref{Fig:150nm} and \ref{Fig:1um}. The nanoparticles in Fig. \ref{Fig:150nm} are smaller than the transition point in Fig. \ref{Fig:stripesize}, and all show the lithiated phase favoring the perimeter. This phenomenon was observed in nanoparticles by Laffont \textit {et al.} \cite{Laffont2006}, who wrote that ``Whatever the samples investigated, ... obtained by either chemical/electrochemical delithiation or by chemical/electro\-chemical lithiation, we always observed from the EELS scan lines spectra particles with \ce{FePO4} as the core and \ce{LiFePO4} as the shell''.

In contrast, the microparticles in Fig. \ref{Fig:1um} and \ref{Fig:2um} lie above the intersection point in Fig. \ref{Fig:stripesize} and show a complex dependence of morphology on size, current, and lithiation vs delithiation. The elastically favorable [101] family of phase boundaries are visible in the interior of many of the particles, and a high frequency of these boundaries was recently observed in electrochemically cycled particles \cite{Mu2016}.

There is a clear morphological asymmetry between charging and discharging in the micron-size particles which can be seen in Figs. \ref{Fig:1um} and \ref{Fig:2um}, particularly at low current. Lithiation (discharge) tends to promote a delithiated core, while delithiation (charge) a lithiated core. This asymmetry has recently been observed by STEM and EELS, where a delithiated/lithiated core was observed during discharge/charge \cite{Honda2015}. Interestingly, this asymmetry is not present at the nano-scale, as can be seen in Fig. \ref{Fig:150nm}.

\begin{figure}[!tp]
\begin{tabular}{ccc}
 \makecell{delithiation\\(charge)}
    & \stackunder{\includegraphics[valign=m,width=.2\columnwidth]{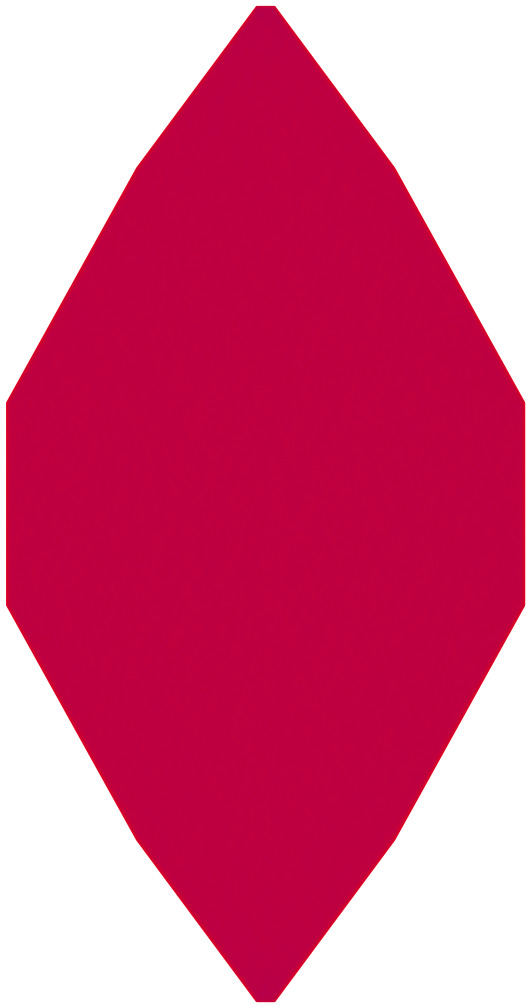}}{X=.75}
      \stackunder{\includegraphics[valign=m,width=.2\columnwidth]{k20-1um-charge.jpg}}{X=.50}
      \stackunder{\includegraphics[valign=m,width=.2\columnwidth]{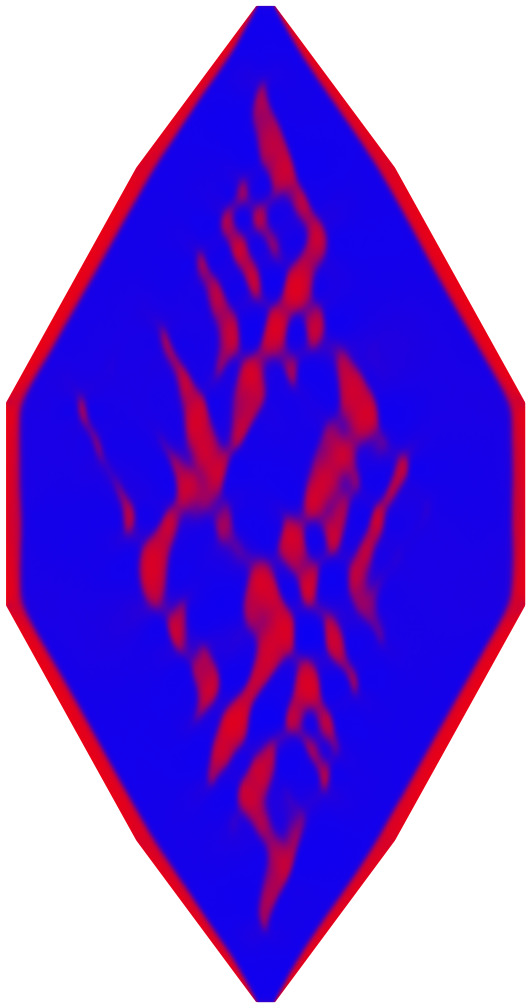}}{X=.25}\\
             \\
 \makecell{lithiation\\(discharge)}
    & \stackunder{\includegraphics[valign=m,width=.2\columnwidth]{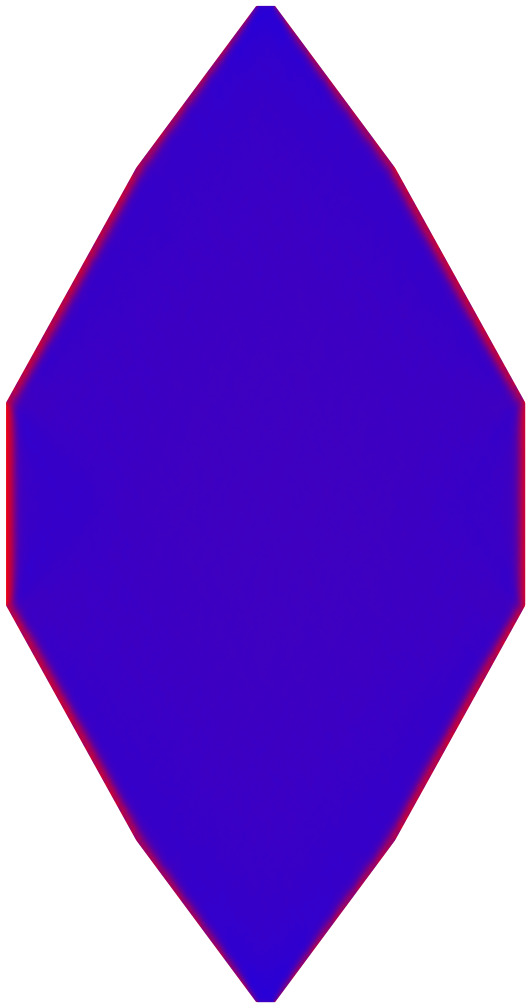}}{X=.25}
      \stackunder{\includegraphics[valign=m,width=.2\columnwidth]{k20-1um-discharge.jpg}}{X=.50}
      \stackunder{\includegraphics[valign=m,width=.2\columnwidth]{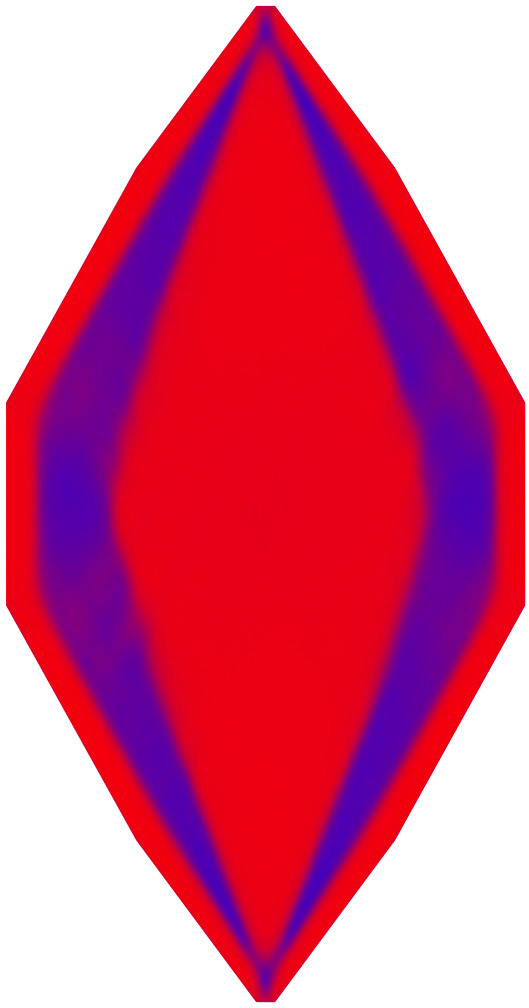}}{X=.75}\\
\end{tabular}
\caption{\unit[1]{um} particles at different states of charge during lithiation/delithiation at a current of $i/i_0=.05$, close to the critical current. As a result of the asymmetric reaction rate, phase separation starts sooner during lithiation but is more heterogeneous during delithiation.}
\label{Fig:asymmetry}
\end{figure}

Another cause of asymmetry is the concentration dependence of the exchange current~\cite{Lim2016}, which modifies the thermodynamic stability of our driven open system~\cite{Bazant2017} and introduces two significant effects, both of which are seen in our simulations in Fig. \ref{Fig:asymmetry}. First, the spinodal points for the onset of linear instability are shifted toward low state of charge, causing phase-separation to begin earlier during lithiation than during delithiation \cite{Liu2014}. Second, the concentration-dependent driven reaction enhances phase separation during delithiation, when the reaction is auto-catalytic, and suppresses phase separation during lithiation, when it is auto-inhibitory, an early prediction of the model~\cite{Bai2011,Cogswell2012,Bazant2013,Bazant2017,Nadkarni2018} confirmed by multiple recent experiments~\cite{Niu2014,Liu2014,Zhang2014,Zhang2015,Lim2016}.

\begin{figure}[!tp]
 \centering
 \subfloat[]{\stackunder{\includegraphics[width=.2\columnwidth]{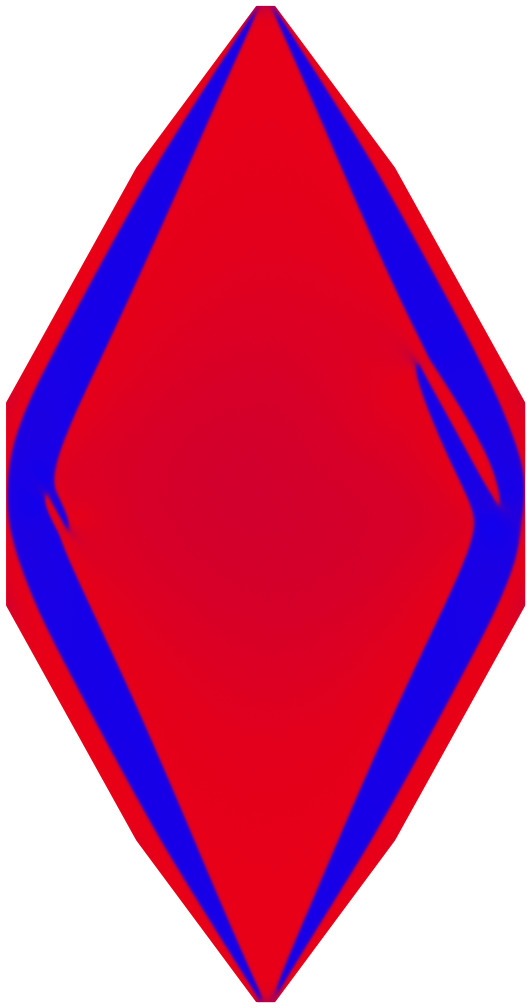}}{X=.7}
             \stackunder{\includegraphics[width=.2\columnwidth]{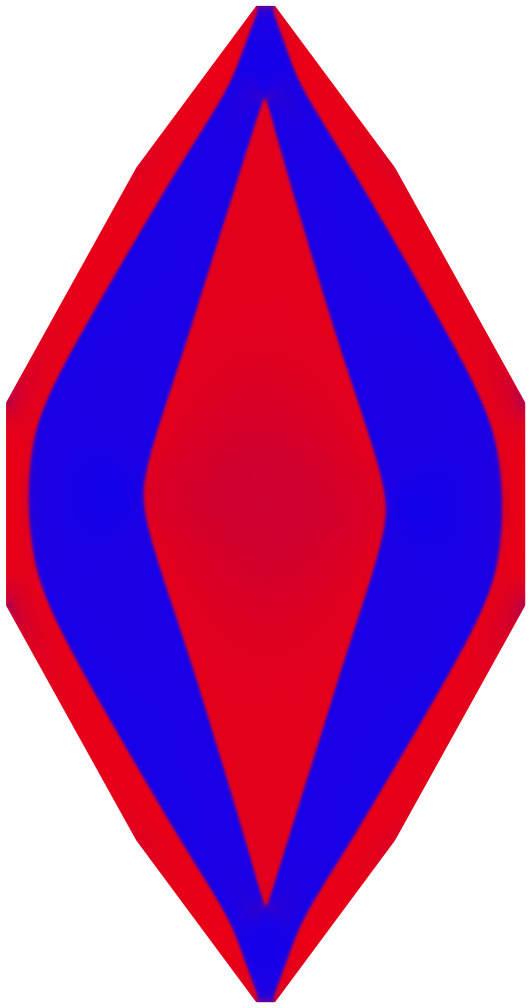}}{X=.5}} 
             \ \ 
 \subfloat[X=.48]{\includegraphics{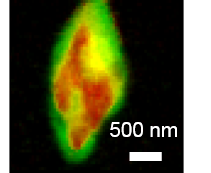}}
             \\
 \subfloat[]{\includegraphics[width=.8\columnwidth]{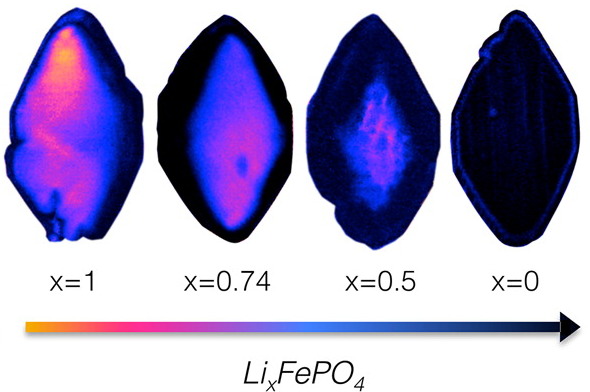}}\\
\caption{Comparison of simulated and observed lithiated diamond cores in \ce{LiFePO4} single particles. (a) Simulated delithiation of a \unit[4]{um} particle with $i/i_0=5\times 10^{-4}$.  (b) A lithiated diamond core in an electrochemically cycled particle that has been allowed to relax for 330 minutes. From \cite{Lim2016}. Reprinted with permission from AAAS.
(c) Chemically delithiated \unit[4]{um} particles imaged using IR near-field spectroscopy. Reprinted with permission from  \cite{Lucas2015}. Copyright 2015 American Chemical Society. }
\label{Fig:diamond_core}
\end{figure}

The lithiated diamond core morphology \cite{Boesenberg2013, Shapiro2014,Nakamura2014,Wang2014,Honda2015,Lucas2015,Yu2015,Wang2016,Lachal2017} is another intriguing feature in micron particles that is captured by our model. Fig. \ref{Fig:diamond_core} shows the striking similarities between the diamond core morphology in simulated, chemically delithiated, and electrochemically delithiated particles. There are two notable features of the diamond morphology. First, the phase boundaries of the diamond align along neither the elastically preferred orientations, nor the external surfaces of the particle (reported in Ref. \cite{Lucas2015}). Second, the simulations indicate that there is a small layer of lithium at the surfaces of the particle with a thickness much less than the stripe period.

This thin wetting layer of lithium present at the side facets is responsible for charge/discharge symmetry at the nanoscale. It was first predicted by \textit{ab initio} calculations \cite{Wang2007}, and later theorized to be responsible for the low size-dependent nucleation barrier in \ce{LiFePO4} \cite{Cogswell2013}. Lithium-rich surfaces may not be widely reported in literature because they represent a small volume of the particle and are difficult to detect experimentally. However, lithium at the side facets has been observed by researchers who specifically looked for it \cite{Laffont2006, Yu2015}.
To investigate whether the symmetry in nanoparticles is controlled by their surfaces, we simulated a \unit[150]{nm} particle with lithium-poor (rather than lithium-rich) surfaces. Charging and discharging remained symmetric, but with a lithiated core. Thus it is the surface properties of the particle that determine the phase morphology at the nanoscale.
 
A key difference between the nano and micro regimes not captured by our model is tendency for fracturing in micron particles \cite{Woodford2010}. Significant fracturing in particles larger than about \unit[1]{um} has been reported \cite{Chen2006,Yu2015}, which almost certainly influences the phase morphology. 
We have observed that defects in the simulated particle tend to relieve strain and attract the lithiated phase. Simulating the effect of fracture would be challenging, but could be done building on model extensions for large deformation mechanics~\cite{DiLeo2014}.

Another neglected effect is surface diffusion, recently identified as an important phenomenon in \ce{LiFePO4}~\cite{Li2018surface}, which counters the effects of driven reactions on thermodynamic stability \cite{Bazant2017}. We previously acknowledged that something like surface diffusion must occur by noting that stripes could only form during chemical delithiation if a particle was somehow able to exchange lithium with itself, but not with neighboring particles \cite{Cogswell2012}. Surface diffusion is currently being incorporated into the model and appears to be the final bit of missing physics necessary to achieve quantitative agreement with \ce{LiFePO4} experiments under conditions far from equilibrium.

\section{Conclusion}
In conclusion, without making any changes to our single-particle intercalation model other than varying the particle size, we have been able to reconcile many of the conflicting reports in literature of the near-equilibrium two-phase morphology in \ce{LiFePO4}. The morphology is largely controlled by the surfaces  in nanoparticles and by elastic strain energy in the micron-sized particles. The transition between the two regimes depends on the particle aspect ratio and occurs when the elastically controlled stripe period becomes smaller than the size of a particle along its a-axis. Complex two-phase morphologies become possible in the micron regime, including asymmetry between charge and discharge. The lithiated diamond core is a common equilibrium morphology that tends to minimize elastic energy in micron-sized particles.

\section*{References}
\bibliographystyle{elsarticle-num}
\biboptions{sort&compress}
\bibliography{micron-particles}
\end{document}